\documentclass[12pt]{article}
\newcommand{\gsim}{\lower.7ex\hbox{$\;\stackrel{\textstyle>}{\sim}\;$}}
\newcommand{\lsim}{\lower.7ex\hbox{$\;\stackrel{\textstyle<}{\sim}\;$}}

\makeatletter
\def\art{\@ifnextchar[{\eart}{\oart}}
\def\eart[#1]#2#3#4#5#6{{\rm #2}, {\em #3 \bf #4} {\rm (#6) #5} ({\em
#1})}
\def\hepart[#1]#2{{\rm #2, \em#1}}
\newcommand{\oart}[5]{{\rm #1}, {\em #2 \bf #3} {\rm (#5) #4}}
\newcounter{alphaequation}[equation]
\def\thealphaequation{\theequation\hbox to
0.6em{\hfil\alph{alphaequation}\hfil}}
\def\eqnsystem#1{
\def\@eqnnum{{\rm (\thealphaequation)}}
\def\@@eqncr{\let\@tempa\relax \ifcase\@eqcnt \def\@tempa{& & &} \or
  \def\@tempa{& &}\or \def\@tempa{&}\fi\@tempa
  \if@eqnsw\@eqnnum\refstepcounter{alphaequation}\fi
\global\@eqnswtrue\global\@eqcnt=0\cr}
\refstepcounter{equation} \let\@currentlabel\theequation \def\@tempb{#1}
\ifx\@tempb\empty\else\label{#1}\fi
\refstepcounter{alphaequation}
\let\@currentlabel\thealphaequation
\global\@eqnswtrue\global\@eqcnt=0 \tabskip\@centering\let\\=\@eqncr
$$\halign to \displaywidth\bgroup \@eqnsel\hskip\@centering
$\displaystyle\tabskip\z@{##}$&\global\@eqcnt\@ne
\hskip2\arraycolsep\hfil${##}$\hfil& \global\@eqcnt\tw@\hskip2\arraycolsep
$\displaystyle\tabskip\z@{##}$\hfil
\tabskip\@centering&\llap{##}\tabskip\z@\cr}

\def\endeqnsystem{\@@eqncr\egroup$$\global\@ignoretrue} \makeatother

\def\vereq#1#2{\lower3pt\vbox{\baselineskip1.5pt \lineskip1.5pt
\ialign{$\m@th#1\hfill##\hfil$\crcr#2\crcr\sim\crcr}}}
\makeatother

\begin{document}

\newcommand{\NP}{Nucl. Phys.}
\newcommand{\PRL}{Phys. Rev. Lett.}
\newcommand{\PL}{Phys. Lett.}
\newcommand{\PR}{Phys. Rev.}

\newcommand{\rem}[1]{{\bf #1}}
\def\Red{}
\def\Black{}
\def\Blue{}

\renewcommand{\thefootnote}{\fnsymbol{footnote}}
\setcounter{footnote}{0}
\begin{titlepage}
\begin{center}

  \hfill     4/8/99\\
	\hfill    LBNL-43091\\
 \hfill    UCB-PTH-99/14\\

\vskip .5in

{\Large \bf \Red
Cosmological Constraints on Theories with Large Extra Dimensions
\footnote
{This work was supported in part by the U.S. 
Department of Energy under Contracts DE-AC03-76SF00098, and in part by the 
National Science Foundation under grant PHY-95-14797 and a graduate 
fellowship. }
}\Black

\vskip .50in

Lawrence J. Hall and David Smith

\vskip 0.2in

{\em Department of Physics and\\ Theoretical Physics Group, Lawrence Berkeley National Laboratory\\
     University of California, Berkeley, California 94720}

\end{center}

\vskip .5in \Blue
\begin{abstract}
In theories with large extra dimensions, constraints from cosmology  
lead to non-trivial lower bounds on the gravitational scale $M$, 
corresponding to upper bounds on the radii of the compact extra 
dimensions.  These constraints are especially relevant to the case of 
two extra dimensions, since only if $M$ is 10 TeV or less do deviations
from the standard gravitational force law become 
evident at distances accessible to planned 
sub-mm gravity experiments.  By examining the graviton decay 
contribution to the cosmic diffuse gamma radiation, we derive, for the 
case of two extra dimensions, a conservative bound $M>110 {\rm TeV}$, 
corresponding to $r_{2}<5.1 
\times 10^{-5} {\rm mm}$, well beyond the reach of these experiments.
We also consider the constraint coming from graviton overclosure of 
the universe and derive an independent bound $M>6.5/\sqrt{h} {\rm 
TeV}$, or $r_{2}<.015h {\rm mm}$. 
\end{abstract}
\end{titlepage}

\renewcommand{\thepage}{\arabic{page}}
\setcounter{page}{1}
\renewcommand{\thefootnote}{\arabic{footnote}}
\setcounter{footnote}{0}

\section{Introduction} 
It was recently proposed that the large hierarchy between the weak and Planck scales 
arises because there exist $n$ extra compact spatial 
dimensions, within which only gravity, and not standard model particles 
and interactions, can propagate\cite{ADDone, ADDtwo, Nima}.  In this 
framework, the Planck scale $M_{P}$ is not a fundamental scale of nature, but 
is rather an effective coupling related to $M$, the scale 
of $(4+n)$ dimensional gravity, by 
\begin{equation}
	M_{P}^{2}=4\pi r_{n}^{n}M^{2+n},
\label{eq:first}
\end{equation}
where $r_{n}$ is the radius of compactification of the $n$ extra 
dimensions\footnote{In this work we assume that the extra dimensions are 
compactified on an $n$ dimensional torus with a single radius.  The 
scale $M$ defined in (\ref{eq:first}) is related to  
Newton's constant in $(4+n)$ dimensions according to $M^{2+n}=(2\pi)^{n}/S_{2+n} 
G^{-1}_{(4+n)}$, where $S_{k}$ is the surface area of a 
unit radius sphere in $k-1$ dimensions.  This is the same definition 
of the gravitational scale used in several recent phenomenological studies 
\cite{{Nima},{peskin},{astro}}.}.  Setting $M \sim {\rm TeV}$ transforms the 
hierarchy problem into the question of why the radii are large.  The approximate values for $r_{n}$ obtained 
for $M \sim {\rm TeV}$ indicate that $n=1$ is ruled out immediately, while for 
the $n=2$ case,  deviations from the standard force law may easily be detected 
by planned experiments sensitive to gravitational forces at distances of 
tens of microns \cite{submm}, depending 
on the precise value of $M$.  
The cases of higher $n$ can be tested instead at high energy colliders\footnote{In 
\cite{Nima} it is shown that if there exist gauge fields that propagate in 
the bulk, they can mediate long range forces relevant to sub-mm experiments, 
regardless of the number of extra dimensions.  In 
this letter we restrict our attention to gravitational forces.}.

Bringing the fundamental scale of gravity down near a TeV dramatically 
alters our view of the universe, and it is not a trivial matter 
that this picture is allowed experimentally.
In \cite{Nima}, a diverse range of collider, astrophysical, and 
cosmological phenomena are examined to verify that the framework is 
in fact safe for all $n>1$.  However, lower bounds on $M_{F}$ from rough
estimates of both energy loss in stellar objects and cosmological 
constraints described in Section 2, cast uncertainty on whether these 
theories can be probed in future sub-mm gravity experiments, even for 
the $n=2$ case.  In this letter we perform a detailed calculation of the most 
stringent cosmological constraints, and derive an upper bound on $r_{2}$ 
that is far below the anticipated range of these experiments.

\section{Cosmology in Theories with Large Extra Dimensions}
In standard cosmology, big bang nucleosynthesis (BBN) provides a 
detailed and accurate understanding 
of the observed light element abundances \cite{BBNrev}.  In order not to lose this 
understanding in the context of theories with large extra dimensions,  we 
must require that before the onset of BBN, the influence of the extra 
dimensions on the expansion  of our 4D wall somehow becomes 
negligible.  In particular we must imagine that starting at some ``normalcy 
temperature'' $T_{*}$, the extra dimensions are virtually empty of energy density
and their radii are fixed.  In \cite{Nima} it is
suggested that the emptiness of the bulk can be explained if $T_{*}$ is the reheat 
temperature following inflation, and if the inflaton is localised on our 4D 
wall and decays only into wall states.  

What is the allowed range for $T_{*}$?  We need $T_{*}>1 {\rm MeV}$ in order 
for ordinary BBN to be recovered.  On the other hand, if $T_{*}$ is too large, then copious
production of bulk gravitons by standard model particles can alter cosmology in unacceptable ways.  
The authors of \cite{Nima} perform rough estimates of several such 
effects and find that the most serious constraints come from overclosure of 
the universe by gravitons and contributions to the cosmic diffuse 
gamma (CDG) radiation from graviton decay.  They estimate that these 
constraints require, for $n=2$, $M \gsim 10{\rm TeV}$, even if the normalcy 
temperature is pushed down to $T_{*}\sim 1 {\rm MeV}$.  As $M$ is 
raised to this level, it becomes unclear whether 
experiments probing macroscopic gravity at small distances will be 
sensitive to the extra dimensions, even if $n=2$.  

In light of the potential implications of cosmological 
constraints on planned experiments, it is worthwhile to calculate 
them more carefully. Detailed studies \cite{BBN1, BBN2} show that, 
in the early universe, the electron neutrinos decouple 
at 1.25 MeV, while the other flavors of neutrinos decouple at 2.15 MeV.  
From the results of  \cite{BBN1}, one can deduce that at  $T={\rm 1MeV}$, the 
relaxation time for muon and tau neutrinos is 10 times longer than the 
inverse Hubble rate of expansion.  If the reheat temperature were less 
than an ${\rm MeV}$, the 
weak interactions would thus be unable to produce the thermal distribution 
of neutrinos required as an initial condition for standard BBN.  For 
this reason we believe that by taking $T_{*}=1 {\rm MeV}$, we suppress the 
cosmological effects of the extra dimensions as much as is conceivably 
allowable, so that bounds we 
derive on $M$ by requiring $T_{*}>1 {\rm MeV}$ should be robust.  We also present bounds 
obtained using the less conservative choice $T_{*}=2.15 {\rm MeV}$, 
which, given that this it is the decoupling temperature for two of the 
three neutrino species, may in fact be a more realistic value.  We find 
that the strongest bounds on $M$ come from the CDG radiation, 
to which we dedicate the bulk of our analysis.

\section{Calculation of the Diffuse Gamma Ray Background}
To calculate the CDG background, we imagine that at the normalcy temperature
$T_{*}$, the bulk is entirely empty, while standard model particles on 
our 4D wall assume thermal distributions.  The 
KK excitations of the graviton are produced through 
the process $\nu \overline{\nu} \rightarrow G$, for example.  The 
spin-summed amplitude squared for this process is\footnote{Feynman rules for the coupling of gravity 
to matter are derived in \cite{{giudice},{Han}}.}
\begin{equation}
\sum |{\cal M}|^2 = \frac{s^2}{4\overline{M}^{2}_{P}},
\end{equation}
where $\overline{M}^{2}_{P}$ is the reduced Planck mass. The number density of 
mass $m$ KK states is then governed by the Boltzmann equation:
\begin{eqnarray*}
\dot{n}_{m}+3n_{m}H &=& \int \frac{d^{3}{\bf p}_{\nu}}{(2\pi)^3 2|{\bf 
p}_{\nu}|} 
\frac{d^{3}{\bf p}_{{\overline\nu}}}{(2\pi)^3 2|{\bf p}_{\overline{\nu}}|} \frac{d^{3}{\bf 
p}_{m}}{(2\pi)^3 2\sqrt{|{\bf 
p}_{m}|^2+m^2}}\\
&&\hspace{1in} \times (2\pi)^4 \delta^4(p_m - p_{\nu} - p_{{\overline \nu}}) 
\sum |{\cal M}|^2 e^{-\frac{|{\bf p}_{\nu}|}{T}} e^{-\frac{|{\bf 
p}_{\overline{\nu}}|}{T}},
\end{eqnarray*}
and the integrations can be performed analytically to obtain
\begin{equation}
s\dot{Y}_m = \dot{n}_{m}+3n_{m}H = \frac{m^5 T}{128\pi^3\overline{M}^{2}_{P}}{\cal 
K}_1(\frac{m}{T}),
\label{eq:prod}
\end{equation}
where ${\cal K}_1$ is a Bessel function of the second kind.  We have 
applied entropy conservation to express the evolution in terms of the 
scaled number density $Y_m=n_{m}/s$, where $s$ is the entropy density.  We will be 
interested in KK states that decay to photons in the MeV range, and from (\ref{eq:prod}) 
we see that essentially all of the graviton production occurs at temperatures 
near $m$ and thus at times well within the radiation dominated era.  The 
neutrino temperature $T$ is therefore
related to the time by \cite{Kolb}  
\begin{equation}
t=1.5 g_{*}^{-1/2} \overline{M}_{P}T^{-2},
\end{equation}
where, since we will be considering temperatures of order MeV and lower, 
$g_{*}=10.75$.  Applying $s \propto T^3$ then leads to a present-day 
graviton density (neglecting decay) of
\begin{equation}
n_{0}^{(m)}=(2.3 \times 10^{-4}) \frac{m 
T_{0}^3}{\overline{M}_{P}}\int^{\infty}_{m/T_{*}}\!dx\, x^3 {\cal K}_1(x),
\label{eq:gravdens}
\end{equation}
where the present day neutrino temperature is $T_{0}=1.96 {\rm K}$.  

A photon produced in the decay of a KK graviton of mass $m$ 
will have a detected energy that depends on the redshift, or 
equivalently, the time, at which the decay occured.  Thus, the energy 
spectrum of photons produced in the decays of mass $m$ KK gravitons can 
be calculated using
\begin{equation}
	\frac{dn_{\gamma}^{(m)}}{dE}=\frac{dn_{\gamma}^{(m)}}{dt}\frac{dt}{dz}\frac{dz}{dE}.
\end{equation}
The derivatives are evaluated by applying $E=\frac{m}{2}(1+z)^{-1}$, 
$t=t_{0}(1+z)^{-3/2}$, and
$n_{\gamma}^{(m)}=2n_{0}^{(m)}\Gamma_{\gamma}/\Gamma_{T}(1-e^{-\Gamma_{T} t}$),  
where $\Gamma_{\gamma}$ is the decay width of the 
graviton into two photons, and $\Gamma_{T}$ is its total decay width.  
We use the time-redshift relation that 
holds for the matter-dominated era, because for KK gravitons that are produced 
near $T_{*}\sim 1{\rm MeV}$, and which decay into photons during the  
radiation-dominated era,  the redshifted photon energies are far below the MeV 
range that interests us.  The spectrum is evaluated to be
\begin{equation}
	\frac{dn_{\gamma}^{(m)}}{dE} = 3n_{0}^{(m)}\Gamma_{\gamma}
	t_{0}(2/m)^{3/2}E^{1/2}e^{-\Gamma_{T}t_{0}(2E/m)^{3/2}}.
\label{eq:spectrum}
\end{equation}
To calculate the full photon spectrum all that remains is to sum over 
KK modes.  This is accomplished using the measure
\begin{equation}
	dN=2 S_{n-1}\frac{\overline{M}_{P}^2}{M^{2+n}}m^{n-1}dm,
\label{eq:measure}
\end{equation}
where 
$S_{n-1}=\frac{2\pi^{n/2}}{\Gamma(n/2)}$ is the surface area of a 
unit-radius sphere in $n$ dimensions.  Using equations (\ref{eq:gravdens}) and 
(\ref{eq:spectrum}), and (\ref{eq:measure}), and the calculated 
width 
\begin{equation}
	\Gamma(G\rightarrow\gamma\gamma)=\frac{m^{3}}{80\pi\overline{M}_{P}^2 },
\label{eq:width}
\end{equation}
we obtain the spectrum
\begin{equation}
	\frac{dn_{\gamma}}{dE}=(1.6 \times10^{-5}) 
	S_{n-1}\frac{t_{0}T_{0}^{3}}{M^{2+n}\overline{M}_{P}}E^{1/2}f_{n}(E,T_{*}),
\label{eq:fullspectrum}
\end{equation}
where the function $f_{n}(E,T_{*})$ is given by
\begin{equation}
	f_{n}(E,T_{*})=\int^{\infty}_{2E}\!dm\, 
		m^{n+3/2}e^{-\Gamma_{T}t_{0}(2E/m)^{3/2}}
		\int^{\infty}_{m/T_{*}}\!dx \,x^{3}
		{\cal K}_{1}(x).
		\label{eq:f}
\end{equation}
Numerically one finds $\Gamma (G \rightarrow 
\gamma \gamma) t_{0} \sim 3\times 10^{-7}(m/{\rm MeV})^{3}$, so that 
for the KK excitations that interest us, the graviton lifetime will 
be much longer that the lifetime of the universe.  Even after 
considering other decay channels, we find that $\Gamma_{T}$ is so 
small that setting the exponential factor in (\ref{eq:f}) to unity does 
not significantly change the values of $f_{n}(E,T_{*})$.

\begin{table}
\begin{center}
\begin{tabular}{|c|c|c|c|c|}\hline
	E(MeV)&$\frac{f_{2}(E,T_{*}=1 {\rm Mev)}}{{\rm 
	MeV}^{9/2}}$&$\alpha_{2}(E)$&$\frac{f_{3}(E,T_{*}=1 
	{\rm Mev})}{({\rm MeV}^{11/2})}$&$\alpha_{3}(E)$\\ \hline
	1&1456&$4.2\times 10^{4}$&9570&$.56$\\
	2&1228&$5.0\times 10^{4}$&8835&$.72$\\
	3&778&$4.0\times 10^{4}$&6571&$.66$\\
	4&379&$2.2\times 10^{4}$&3801&$.44$\\
	5&150&$9.8\times 10^{3}$&1773&$.22$\\
	6&51.1&$3.6\times 10^{3}$&698&$1.0 \times 10^{-1}$\\
	7&15.5&$1.2\times 10^{3}$&240&$3.6 \times 10^{-2}$\\
	8&4.27&$3.6\times 10^{2}$&74&$1.2 \times 10^{-2}$\\
	9&1.09&94&21.2&$3.6 \times 10^{-3}$\\
	10&.263&24&5.6&$1.0 \times 10^{-3}$\\ \hline
\end{tabular}
\caption{Values of the parameters $\alpha_{n}(E)$ and 
$f_{n}(E,T_{*}=1{\rm MeV})$ defined in equations (\ref{eq:f}) and(\ref{eq:alpha}).}
\label{table:table}
\end{center}
\end{table}

Taking $t_{0}=10^{10}$ years, we find that for $T_{*}=1 {\rm MeV}$, the spectrum can be written as
\begin{eqnarray}
     \left.\frac{dn_{\gamma}}{dE}\right|_{T_{*}=1{\rm MeV}}& =& 4.6 \times 
     10^{-6(n-2)}S_{n-1}\frac{f_{n}(E,T_{*}=1 {\rm MeV})}{{\rm 
     MeV}^{(n+5/2)}}\nonumber \\
	 &&\hspace{.5in}\times \left(\frac{E}{{\rm 
	 MeV}}\right)^{1/2}\left(\frac{M}{{\rm TeV}}\right)^{-(n+2)} 
     {\rm MeV}^{-1}{\rm cm}^{-2}{\rm s}^{-1}{\rm ster}^{-1}\\
	 &\equiv&\alpha_{n}(E)\left(\frac{M}{{\rm TeV}}\right)^{-(n+2)}{\rm 
	 MeV}^{-1}{\rm cm}^{-2}{\rm s}^{-1}{\rm ster}^{-1}.
\label{eq:alpha}
\end{eqnarray}
Values for $\alpha_{n}(E)$ and $f_{n}(E,T_{*}=1{\rm MeV})$ for $n=2,3$ 
are given in Table \ref{table:table}.

The above photon spectrum was derived by 
calculating the density of KK gravitons produced by  annihilation of a single neutrino species.  Repeating 
the same calculation for $\gamma \gamma$ annihilation, we find a 
spin-summed amplitude squared
\begin{equation}
	\sum |{\cal M}|^2 = 2 \frac{s^2}{\overline{M}^{2}_{P}},
\end{equation}
and, taking into account the symmetry factor of $1/2$ due to the initial 
state photons, a contribution to the spectrum that is larger than that 
coming from a single neutrino flavor by a factor of 4.  When 
comparing with the observed spectrum, we will take the sum of 
contributions from photons and three flavors of neutrinos\footnote{We 
neglect an additional contribution from $e^{+}e^{-}$ annihilation for 
the sake of a simplified calculation.  Including this
contribution enhances the bounds we derive only slightly.}:
\begin{equation}
	\left.\frac{dn_{\gamma}}{dE}\right|_{T_{*}=1{\rm 
	MeV}}=7\alpha_{n}(E)\left(\frac{M}{{\rm TeV}}\right)^{-(n+2)}{\rm 
	 MeV}^{-1}{\rm cm}^{-2}{\rm s}^{-1}{\rm ster}^{-1}.
\label{eq:calc}
\end{equation}

Before comparing our results with the CDG background data, we can 
already obtain an independent bound on $M$ by requiring that the KK 
gravitons do not overclose the universe.  Contributions 
from photon and neutrino annihilation give a graviton energy 
density
\begin{equation}
	\rho_{G}=14S_{n-1}\frac{\overline{M}_{P}^2}{M^{2+n}}\int^{\infty}_{0}\!dm\, m^{n}n_{0}^{(m)},
\end{equation}
 where $n_{0}^{(m)}$ is the density defined in (\ref{eq:gravdens}).  For 
 $n=2$ we obtain 
 \begin{equation}
 	\rho_{G}=14\times 10^{-44} \left(\frac{M}{{\rm TeV}}\right)^{-4} {\rm GeV}^{4},
 \end{equation}
which, upon comparison with $\rho_{c}=8.1 h^{2} 10^{-47} {\rm 
GeV}^{4}$, 
leads to $M>6.5/\sqrt{h}{\rm TeV}$.
Using the relation between the fundamental scale and the radius of 
compactification,
\begin{equation}
	r_{n}=2\times 10^{31/n - 16}\left(\frac{1 {\rm TeV}}{M}\right)^{1+2/n}{\rm mm},
\label{eq:radius}
\end{equation}
we obtain
\begin{equation}
	r_{2}<.015h {\rm mm}.
\label{eq:critdens}
\end{equation}
It may be possible, although certainly challenging, to probe distances of this size in near-future sub-mm gravity
experiments.  If we take a less conservative bound on $T_{*}$ and 
instead use $T_{*}=2.15 {\rm MeV}$, the decoupling temperature for the 
muon and tau neutrinos, we get the more stringent bounds $M>13.9/\sqrt{h}{\rm TeV}$ 
and $r_{2}<3.3h\times 10^{-3} {\rm mm}$.  Distances this small are 
likely not to be accessible to those experiments.
\section{Comparison with Data}
The CDG background has been measured recently in the 800 keV to 30 MeV 
energy range using the COMPTEL instrument\cite{kappadath}.  The 
authors of \cite{kappadath} find that the photon spectrum is well described by the 
power-law function $A(E/E_{0})^{-a}$, with $a=-2.4 \pm .2$, 
$E_{0}=5{\rm MeV}$, and $A=(1.05 \pm 0.2) \times 10^{-4}{\rm MeV}^{-1}{\rm 
cm}^{-2}{\rm s}^{-1}{\rm ster}^{-1}$.  They find no evidence for the 
``MeV bump'' that was inferred from previous data.  Using the COMPTEL 
results and the calculated contribution to the 
background from graviton decay in equation (\ref{eq:calc}), we can place a lower 
bound on the gravitational scale $M$:
\begin{equation}
	\left( \frac{M}{{\rm TeV}}\right)^{n+2} > 7\alpha_{n}(E)\left( 
	\frac{\left.\frac{dn_{\gamma}}{dE}\right|_{measured}}{{\rm 
	 MeV}^{-1}{\rm cm}^{-2}{\rm s}^{-1}{\rm ster}^{-1}} \right)^{-1}.
\end{equation}
We find that the most stringent bounds are obtained for $E \simeq 
4{\rm MeV}$.  Using the very conservative upperbound 
$\left.\frac{dn_{\gamma}}{dE}\right|_{measured} < 10^{-3}{\rm MeV}^{-1}{\rm cm}^{-2}
{\rm s}^{-1}{\rm ster}^{-1}$ gives, for $n=2$, 
\begin{equation}
M > 110 {\rm TeV}.  
\label{eq:mbound}
\end{equation}
This corresponds to a bound on the radius of compactification of 
\begin{equation}
	r_{2}< 5.1 \times 10^{-5} {\rm mm},
\end{equation}
which is far smaller than the distances at which gravity can be 
probed in planned experiments.  If we instead use $T_{*}=2.15 {\rm MeV}$, we obtain 
$M > 350 {\rm TeV}$: $M$ must be about $10^{3}$ or more larger 
than the electroweak VEV, reintroducing a mild hierarchy problem, and 
hence requiring supersymmetry or some other solution\footnote{The string scale may be lower than $M$, in 
which case the hierarchy is alleviated slightly.  At least in the 
string scenario described in \cite{ADDtwo}, where standard model 
particles are localized on a 3-brane, 
the factor one might gain in this way is $\sim 10$ 
rather than $\sim 10^{3}$\cite{Nima}. If the standard model particles 
are instead localized on a brane of higher dimension, one can achieve 
further suppression of the string scale relative to $M$\cite{shiu}.}.  Applying the same experimental bound 
to the $n=3$ case leads to $M > 5.0{\rm TeV}$ or $M > 13.8 {\rm 
TeV}$, for $T_{*}=1 {\rm MeV}$ and $T_{*}=2.15 {\rm MeV}$, respectively.
\section{Cosmological Uncertainties}
Are there ways to evade our bounds on $M$?  
The authors of \cite{Nima} have pointed out that there may be 
additional branes, besides our own, on which gravitons can decay.  
Depending on the decay rate on these branes, their existence can greatly reduce the 
number of gravitons that decay on our brane.  If $1/\Gamma'$, the 
decay lifetime onto the other brane(s), is significantly 
longer than the age of 
the universe $t_{0}$, then the number of decays on our brane will not 
be substantially reduced.  If $1/\Gamma' \ll t_{0}$, on the other 
hand, the number of decays on our brane, and thus the contribution to 
the photon background, is reduced by a factor $\sim 1/(\Gamma' t_{0})$.  Moreover, in this case 
nearly all of the gravitons decay at large redshift, so that for $T_{*} 
\sim 1 {\rm MeV}$ the redshifted photon energies fall below 
the ${\rm MeV}$ range.

We know of two scenarios that give the large $\Gamma'$ required to
evade the CDG bound.  In the first, $\Gamma'$ is large because the 
extra brane(s) have higher dimension than ours \cite{Nima}.  If 
one of these so called ``fat-branes'' has thickness $W$ in a single extra 
dimension, the probability that a graviton will decay on it is enhanced over its probability of 
decaying on our brane by a factor $\sim WT_{*}$.  For $WT_{*}\sim 
5\times 10^{6}$, we find that the graviton 
contribution to the CDG is consistent with the COMPTEL result for  
$M$ as low as $\sim 1{\rm TeV}$.  Taking $T_{*}=1{\rm MeV}$, this corresponds 
to a thickness $W>1 \mu {\rm m}$.  Note that 
introducing a higher-dimensional brane does {\em not} enable us to evade the bound obtained 
by considering overclosure of
the universe, equation (\ref{eq:critdens}).  Because the fat-brane is higher-dimensional,
the decay products 
have a momentum component that is perpendicular to our brane, and 
which therefore does not redshift (recall that the extra spatial 
dimensions are frozen).  Thus the energy density of these decay 
products will go as $R^{-3}$ rather than $R^{-4}$, regardless of 
whether or not the particles are relativistic, and we cannot eliminate 
the graviton contribution to $\Omega$. 

In the second scenario, $\Gamma'$ is large because there exist a very large number
of 4D branes in addition to our own.  More precisely, we need at 
least $\sim 5\times 10^{6}$ additional branes to have a graviton contribution to the 
CDG background that is consistent with the COMPTEL result when 
$M \sim 1{\rm TeV}$.  An important distinction between this 
scenario and the one involving higher dimensional branes is that now,
provided the foreign branes are parallel to our own, 
relativistic decay products on 
them {\em do} redshift, and the bound in equation 
(\ref{eq:critdens}) can be evaded.

\section{Conclusions}
We have examined two cosmological constraints on the theories with large 
extra dimensions proposed in \cite{ADDone, ADDtwo, Nima}.  To place 
limits on $M$, we apply a conservative lower bound on the normalcy 
temperature, $T_{*}>1{\rm MeV}$, as required by BBN.  We find that, ignoring the possibile
existence of additional branes, the radius of compactification of the 
extra dimensions for $n=2$ is bound by the cosmic diffuse gamma ray background, 
to be $r_{2}< 5.1 \times 10^{-5} {\rm mm}$, 
well beyond the reach of planned sub-mm gravity experiments.  From the constraint that gravitons do not overclose 
the universe we derive a milder bound, $r_{2}<.015h {\rm mm}$, albeit one 
that is less dependent on our assumptions regarding foreign branes.  If 
one instead insists on a normalcy temperature above the decoupling 
temperature for the muon and tau neutrinos, $T_{*}>2.15 {\rm MeV}$, these bounds become $r_{2}< 
5.2 \times 10^{-6} {\rm mm}$ and $r_{2}<3.3h \times 10^{-3}{\rm mm}$, 
respectively.

A recent calculation has given the bound $M>50 {\rm TeV}$ for 
$n=2$, from the requirement that supernovae do not cool too rapidly 
by graviton emission \cite{astro}.  This astrophysical constraint 
complements the cosmological ones we have studied:  it is subject to 
larger technical calculational uncertainties, while our analysis is 
subject to uncertainties in the global cosmological picture.  In 
either case, a bound on $M$ can only be translated into a limit on
$r_{n}$ if it is assumed that the extra dimensions have the same 
size.  No matter how large $n$ is taken to be, it is always possible 
that one extra dimension has a size in the mm - $\mu$m range, while 
the others are much smaller\cite{dvali}.  However, in a framework involving 
vastly different radii, we are unable to argue why gravity would be expected 
to diverge from $r^{-2}$ behavior 
specifically at those distance scales accessible to planned experiments.
\\

\noindent {\em Acknowledgments}:  We are grateful to Nima Arkani-Hamed for 
useful discussions.  D.S. also thanks Michael Graesser for helpful 
conversations.

\end{document}